\documentclass[10pt,twocolumn,twoside,letterpaper]{IEEEtran}
\usepackage{cite}
\usepackage{amsmath,amssymb,amsfonts}
\usepackage{graphicx}
\usepackage{textcomp,nicefrac}
\usepackage{caption,subcaption}

\usepackage{geometry}
\usepackage{url}

\begin{document}
\title{Performance of the new Readout Electronics for the ATLAS (s)MDT Chambers and Future Colliders at High Background Rates
}
%
%
%

\author{G. H. Eberwein, 
	   O. Kortner, 
	   H. Kroha,
	   E. Voevodina

\thanks{G. H. Eberwein was with Max-Planck-Institute for Physics, Foehringer Ring 6, 80805 Munich, Germany. He is now with Department of Physics, University of Oxford, Denys Wilkinson Building, Keble Road, Oxford OX1 3RH, UK (e-mail: gregor.eberwein@physics.ox.ac.uk).}
\thanks{O. Kortner, H. Kroha, E. Voevodina are with Max-Planck-Institute for Physics, Foehringer Ring 6, 80805 Munich, Germany.}}

\maketitle

\pagenumbering{gobble}

\begin{abstract}
Small-diameter Drift Tube (sMDT) detectors with 15~mm tube diameter have proven to be excellent candidates for precision muon tracking detectors in experiments at future hadron colliders like HL-LHC and FCC-hh where unprecedentedly high background rate capabilities are required. sMDT chambers are currently being installed in the inner barrel layer of the ATLAS muon spectrometer. The rate capability of the sMDT drift tubes in terms of muon detection efficiency and spatial resolution is limited by the performance of the readout electronics. A new (s)MDT ASD (Amplifier-Shaper-Discriminator) readout chip for use at the HL-LHC and future hadron colliders with a faster peaking time compared to the old chip has been developed, reducing the discriminator threshold crossing time jitter and thus improving the time- and spatial resolution with and without $\gamma$-background radiation. Additionally, a method compensating the gas gain drop due to space charge at high $\gamma$-background hit flux by adjusting the sMDT operating voltage will be presented. Simulations show, that the addition of active baseline restoration circuits in the front-end electronics chips in order to suppress signal-pile-up effects at high counting rates further leads to significant improvement of both efficiency and resolution. Extensive tests using sMDT test chambers have been performed at the CERN Gamma Irradiation Facility (GIF++). Chambers equipped with new readout chips with improved pulse shaping and discrete readout circuits with baseline restoring functionality have been tested.
\end{abstract}

\section{Introduction}
\label{sec:introduction}
\IEEEPARstart{I}{n} the next two decades, hadron collisions at the future high luminosity LHC (HL-LHC) with an expected instantaneous luminosity of $\mathcal{L}=7.5\cdot10^{34}\, \mathrm{cm^{-2}s^{-1}}$, will provide new opportunities for searches for physics beyond the Standard Model. The achievable centre-of-mass energy of the HL-LHC is $\sqrt{s}=14\,\mathrm{TeV}$. To overcome this limitation, a successor to the HL-LHC has been proposed - the Future Circular Collider (FCC-hh) with a circumference of 100 km and a centre-of-mass energy of $\sqrt{s}=100\,\mathrm{TeV}$ for proton collisions, opening up a new unexplored regime for high energy particle physics. An instantaneous luminosity of up to $\mathcal{L}=30\cdot10^{34}\, \mathrm{cm^{-2}s^{-1}}$ is foreseen.

The muon detector system has two important functions: Precise muon track reconstruction with high spatial resolution and fast timing information for triggering on potentially interesting collision events. Monitored Drift Tube detectors (MDT) are used in the ATLAS muon spectrometer for precise muon track reconstruction. Small-diameter Monitored Drift Tube (sMDT) detectors have been developed to cope with the increased background counting rates expected at the HL-LHC and FCC-hh, scaling with the instantaneous luminosity. To support the new (s)MDT based high-resolution trigger system with continuous readout of the muon chambers and the increased overall trigger rates and further enhance the rate capability of sMDT detectors, a new (s)MDT readout chip has been developed which has been characterized with and without background irradiation. In addition, methods have been investigated to increase the rate capability even further, in view of the application at future hadron colliders like the FCC-hh where a $\gamma$-background hit flux up to 25~kHz/cm\textsuperscript{2} is expected in the inner end-cap and forward muon system. This corresponds to a background counting rate of 1.25~MHz/tube for sMDT detectors with 0.4~m tube length. In the barrel and outer end-cap muon system of the FCC-hh baseline reference detector up to 1.25~kHz/cm\textsuperscript{2} background hit flux and a background counting rate of 500~kHz/tube for 2.8~m long sMDT tubes is expected. These rate values include a safety factor of 2.5. Proposed sMDT chambers consist of 8 tube layers, mounted together as two multilayers of four each, and being spaced apart 1.5~m by a spacer frame. To achieve the goal of 70~$\mu$rad angular resolution, a single tube spatial resolution of minimum $150\mu$m is required, while ensuring a 99\% track reconstruction efficiency requiring at least four hit layers of the sMDT chamber.

\section{sMDT Drift Tube Detectors and High Rate Effects}
\label{sec:high_effects}
Aluminium sMDT drift tubes have a diameter of 15~mm and a wall thickness of 0.4~mm. They are filled with an Ar/CO\textsubscript{2} (93/7) gas mixture at 3~bar absolute and are operated at 2730~V resulting in a nominal gas amplification factor $G$ of $2\cdot10^{4}$.

\begin{figure}[t]
\centerline{\includegraphics[width=2.9in]{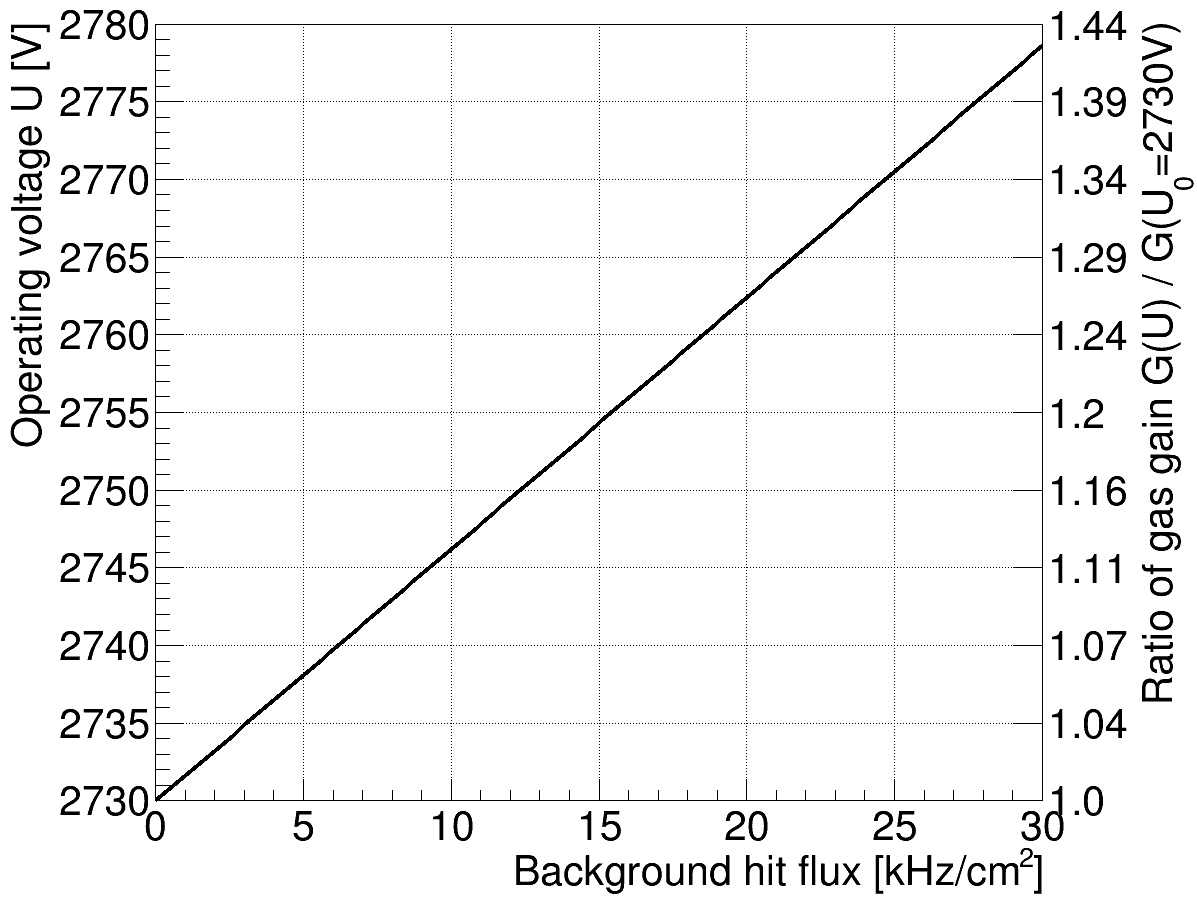}}
\caption{Calibration curve for operation at adjusted operating voltage to compensate the gain loss encountered at high $\gamma$-background hit fluxes. The ratio of gas gain at adjusted operating voltage over nominal gas gain is calculated from the Diethorn formula and shown on the right axis.}
\label{CalibCurve}
\end{figure}

The degradation of spatial resolution and detection efficiency in sMDT detectors at high background hit rates is caused by three effects. A background hit can mask a muon hit with increasing probability at high background rates. The multilayer structure of sMDT detectors comprising 8 tube layers provides redundancy allowing for operation up to 30\% occupancy, minimizing this effect. Slowly drifting positive ions lead to space charge in the drift tubes, increasing with the background hit flux. Induced space charge reduces the gas gain, leading to a decrease in spatial resolution on the order of 20\% at 25~kHz/cm\textsuperscript{2} background hit flux, whereas the detection efficiency is not affected strongly due to the operation of the readout electronics at a high gain. A new method is presented, compensating the gas gain loss at high background rates by adjusting the operating voltage according to the $\gamma$-background hit flux encountered. This is viable due to the linearity of the Ar/CO\textsubscript{2} (93/7) drift gas space-to-drift time relation for the maximum drift radius of 7.1~mm for sMDT tubes, eliminating the effect of space charge fluctuations. A first order calibration curve for the operating voltage as a function of the background hit flux and the corresponding increase in nominal gas gain based on the Diethorn formula is shown in Figure \ref{CalibCurve}. 
The Diethorn formula models in good approximation the gas gain $G$ in absence of space charge: 

\begin{equation}
ln\,G=ln\,2\,\,\frac{r_{min}E(r_{min})}{\bigtriangleup V}\,ln\left(\frac{E(r_{min})}{E_{min}(\rho_0)\frac{\rho}{\rho_0}} \right) \, ,
\label{eqtn:Diethorn}
\end{equation}

\noindent where $E(r_{min})$ is the electric field at the anode wire, and $\rho_0$ and $\rho$ are the normal and actual gas density, respectively. $E_{min}(\rho_0)$ is the electric field required to start an avalanche in the given gas and $\bigtriangleup V$ the ionisation potential. The adjusted operating voltage $U_0'(\Phi)$ in dependence of the background hit flux $\Phi(\nu) [kHz/cm^{2}]$ is calculated as:

\begin{equation}
U_0'(\Phi)=U_{0}+\frac{Q_p G(U_0)\, \bigtriangleup \! t_{ion}\, r_{max}}{2\pi^2\epsilon_0}\, \Phi(\nu)\, ,
\label{eqtn:Voltage_adjustment}
\end{equation}

\noindent where $\bigtriangleup t_{ion}$ is the approximated ion drift time of 1.02~ms for a maximum drift radius $r_{max}$ of 7.1~mm. $Q_p G(U_0)= 3.91\pm0.05$~pC represents the measured average charge deposited by a background hit after multiplication with the nominal gas gain $G(U_0=2730V)$ for the environment at the Gamma Irradiation Facility (GIF++) at CERN.

\begin{figure}[th]
\centerline{\includegraphics[width=2.9in]{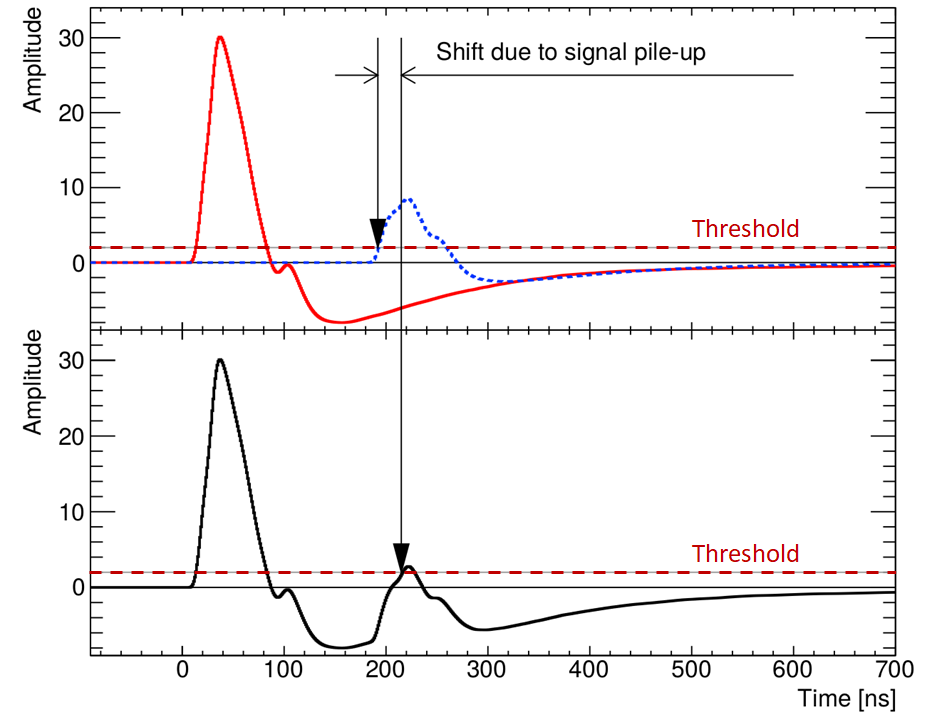}}
\caption{Effect of signal pile-up for bipolar shaping of the readout chip.}
\label{PileUp}
\end{figure}

The third effect is caused by the signal pile-up in the readout electronics. Bipolar shaping with the characteristic undershoot of equal area after the positive signal pulse is used in order to avoid baseline shift associated with unipolar shaping. When a signal generated by a background hit (red pulse in Figure \ref{PileUp}) is succeeded by a muon hit (blue pulse), the muon signal is superimposed on the undershoot of the preceding background pulse (black curve) with increasing probability for increasing background hit rate. As the pulses have finite rise time, the threshold crossing is delayed for the superimposed signal, leading to an additional jitter in the measurement of the drift time and thus the drift radius. The consequence is a deterioration of spatial resolution. Additionally, a muon hit superimposed on the undershoot can be suppressed altogether, reducing the muon detection efficiency.
Two solutions are proposed: 1.~Faster peaking time as for the new readout chip and 2.~Active baseline restoration, aiming to suppress the undershoot caused by bipolar shaping altogether.

\section{New Readout Chip Performance}

\begin{figure}[b]
\centerline{\includegraphics[width=2.9in]{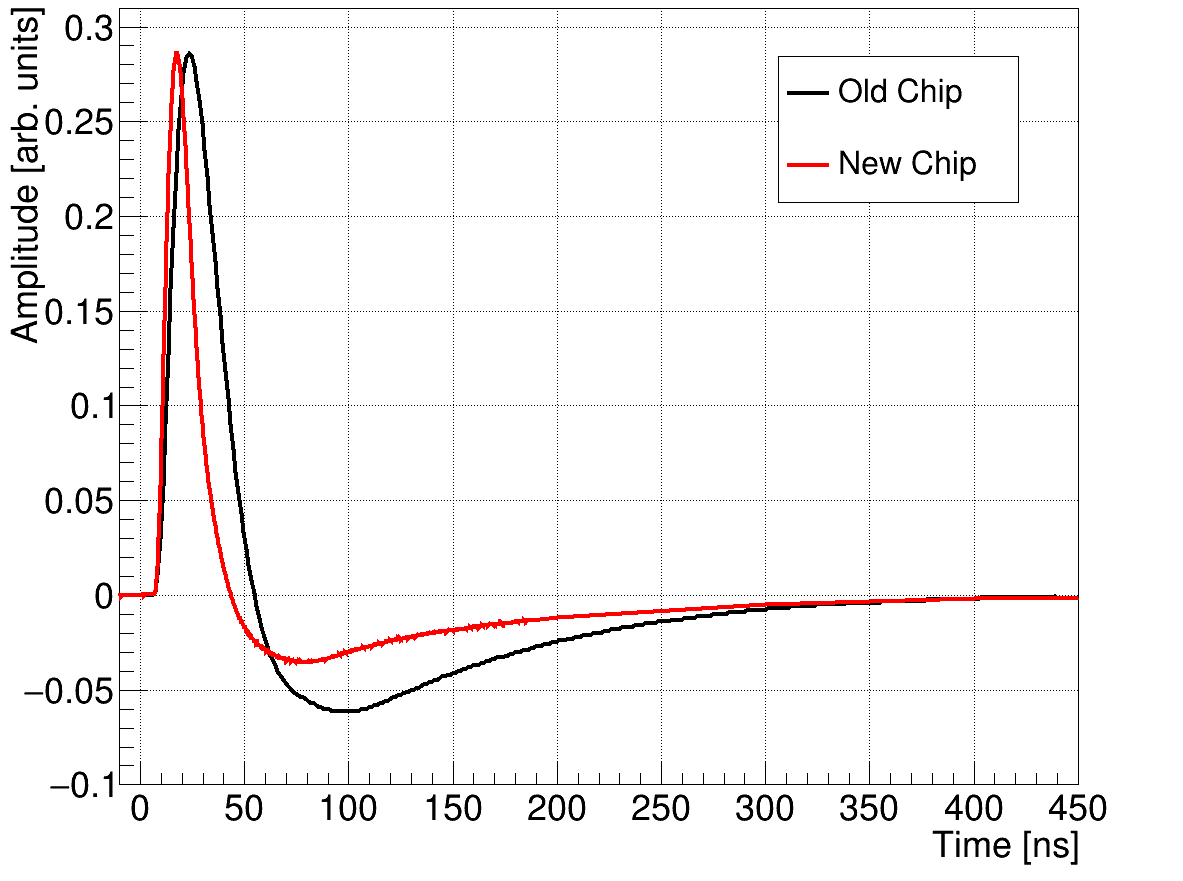}}
\caption{Delta pulse response of the old (ASD-0) and new (ASD-6) readout chip, illustrating the reduced peaking time.}
\label{PeakTime}
\end{figure}

A new ASD (Amplifier-Shaper-Discriminator) readout chip (ASD-6) with 3~ns faster peaking time compared to the old chip (ASD-0) has been developed (see Figure \ref{PeakTime}), reducing discriminator threshold crossing time jitter and thus improving the sMDT tube spatial resolution on the order of 20\% with and without background irradiation. Measurements have been performed at the Gamma Irradiation Facility (GIF++) at CERN, which houses an intense \textsuperscript{137}Cs source with 12~TBq activity, emitting 662~keV photons. The photon background hit flux is adjusted with a set of remote controlled attenuation filters. Cosmic muon tracks have been used in absence of a testbeam during data taking, requiring a multiple scattering (MS) correction on the measured spatial resolution, which has been simulated in GEANT-4. The MS contribution of $88\pm5$~$\mu$m, independent of the $\gamma$-background hit flux, has been subtracted quadratically from the spatial resolution measurement. Two test chambers were installed in the test setup: 1.~sMDT chamber with 23~cm tube length, read out by the old ASD-0 readout chip and 2.~sMDT chamber with 160~cm tube length, read out by the new ASD-6 readout chip.


\begin{figure}[t]
\centerline{\includegraphics[width=2.9in]{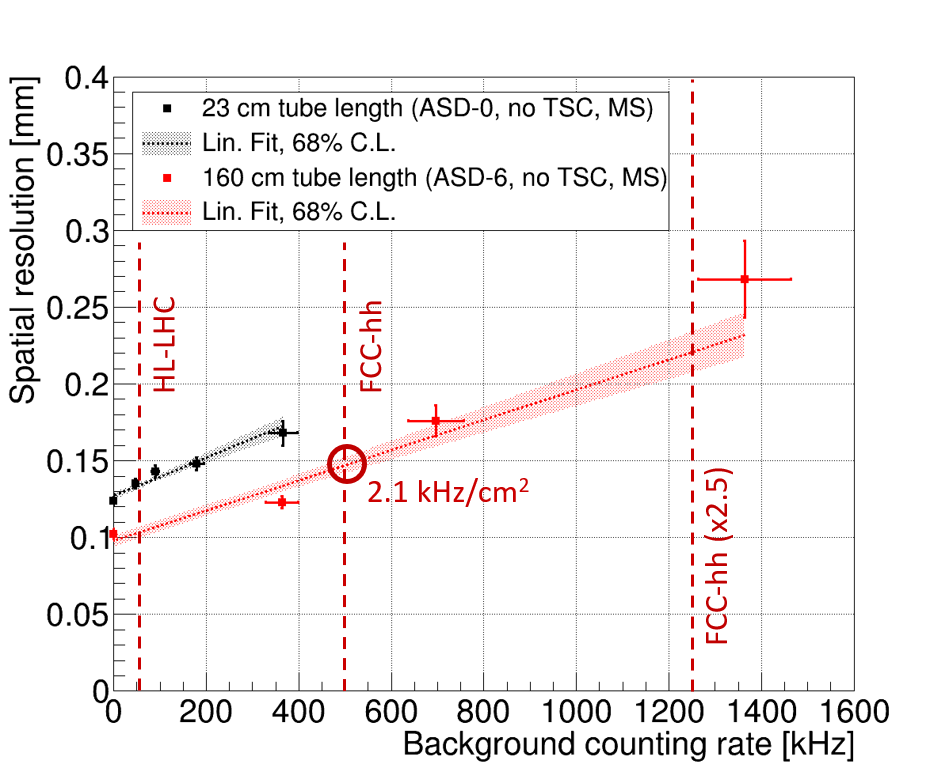}}
\caption{Improved average spatial resolution of the new readout chip as a function of the $\gamma$-background hit rate for sMDT drift tubes. Maximum upper limits for operation at HL-LHC and FCC-hh conditions are indicated as dashed lines.}
\label{ResRate}
\end{figure}

Figure \ref{ResRate} shows the improved average spatial resolution for the new readout chip (ASD-6) compared to the old readout chip (ASD-0) as a function of different background hit rates. the goal of 150~$\mu$m spatial resolution is achieved for the 160~cm long sMDT tubes read out with the new ASD-6 readout chip at a $\gamma$-background hit flux of 2.1~kHz/cm\textsuperscript{2}, corresponding to 500~kHz/tube background counting rate. This ensures safe operation in FCC-hh conditions in the barrel and outer end-cap region for the proposed muon spectrometer.

\begin{figure}[t]
\centerline{\includegraphics[width=2.9in]{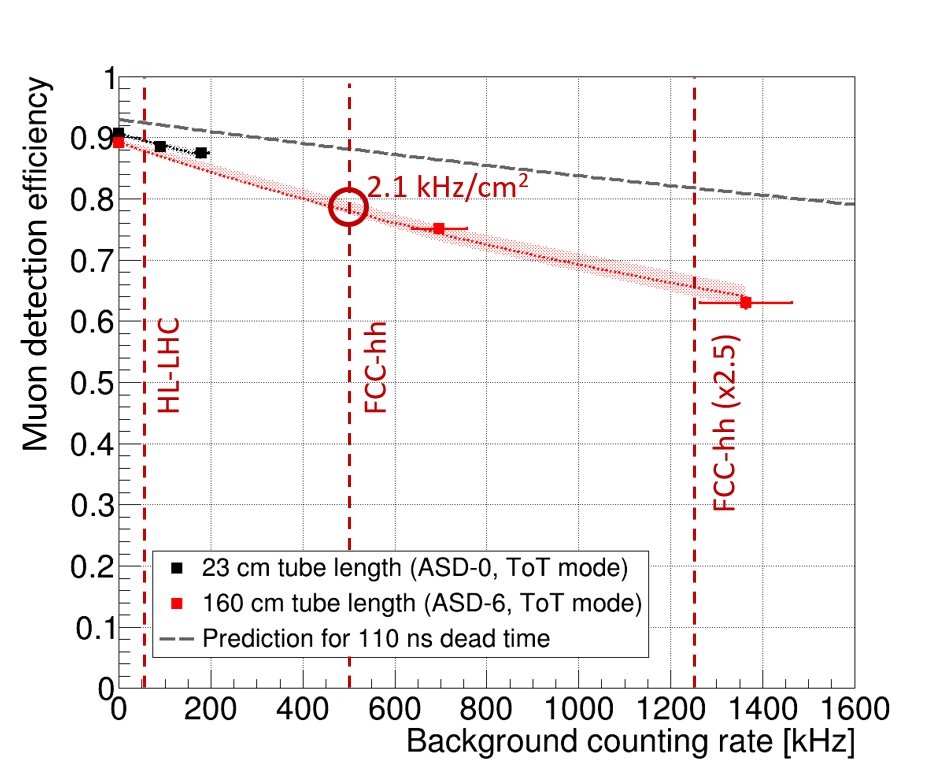}}
\caption{Muon detection efficiency in dependence of the $\gamma$-background counting rate. Maximum upper limits for operation at the HL-LHC and FCC-hh are indicated as vertical dashed lines.}
\label{Efficiency}
\end{figure}

The muon detection efficiency for the old and new readout chip is shown in Figure~\ref{Efficiency}. The dashed line indicates the minimum efficiency loss caused by the $\gamma$-background hit induced electronics dead time (on the order of 110~ns) in dependence of the $\gamma$-background counting rate. The electronics signal pile-up, given the bipolar shaping technique, suppresses muon hits with increasing probability at higher rates, leading to an effective electronics dead time on the order of 300~ns. This difference in dead time and therefore muon detection efficiency illustrates the potential of implementing active baseline restoration circuitry. An average tube muon detection efficiency of 80~\% is sufficient to achieve a 99~\% track reconstruction efficiency when requiring four hits in 8 tube layers of a sMDT chamber. The goal for operation in FCC-hh conditions is therefore achieved for background counting rates up to 500~kHz/tube as expected in the  barrel and outer end-cap muon systems.

Improvements in both spatial resolution and muon detection efficiency are required for reliable operation of sMDT chambers in the inner end-cap and forward region of the proposed FCC-hh muon spectrometer.

\section{Further Improvements of Spatial Resolution}
\subsection{Operating Voltage Adjustment}

\begin{figure}[b]
\centerline{\includegraphics[width=2.9in]{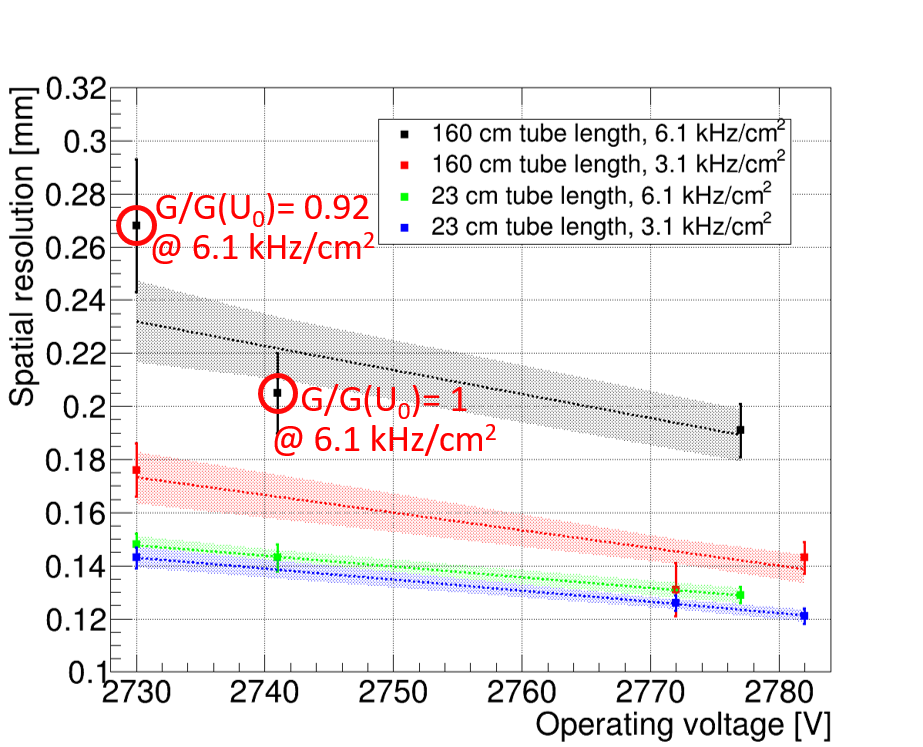}}
\caption{Improvement of the average drift tube spatial resolution with increasing operating voltage $U$ for two background hit fluxes in the two sMDT test chambers (23~cm and 160~cm tube length).
 Linear fits with 68\% confidence interval error bands are shown. The reduction of the nominal gas gain due to induced space charge, and correction, is marked exemplary in red for the 160~cm long sMDT tubes at a background hit flux of 6.1~kHz/cm\textsuperscript{2}.}
\label{HV_res}
\end{figure}

The effect of operating voltage adjustment in dependence of the background hit flux on the spatial resolution, as discussed in Section~\ref{sec:high_effects}, is illustrated in Figure~\ref{HV_res}. The operating voltage is increased such, that the induced space charge decreases the gas gain back to the nominal value of $2\cdot10^{4}$. Measurements have been performed at  background hit fluxes of 3.1~kHz/cm\textsuperscript{2} and 6.1~kHz/cm\textsuperscript{2} for both sMDT test chambers. In case of the 160~cm long sMDT tubes at 6.1~kHz/cm\textsuperscript{2} background hit flux, the nominal gas gain is reduced to 92~\% by $\gamma$-background hit induced space charge. To compensate for the loss in spatial resolution, the operating voltage is increased to 2740~V. Now the induced space charge reduces the gas gain exactly to the nominal level of $2\cdot10^{4}$, and leads to an improvement in spatial resolution on the order of 5~\%. The gain drop effect on the spatial resolution increases with the background hit flux, and measurements with higher background hit fluxes up to 25~kHz/cm\textsuperscript{2} will be performed to confirm these results.

\subsection{Active Baseline Restoration}

\begin{figure}[h]
\centering
\includegraphics[width=2.9in]{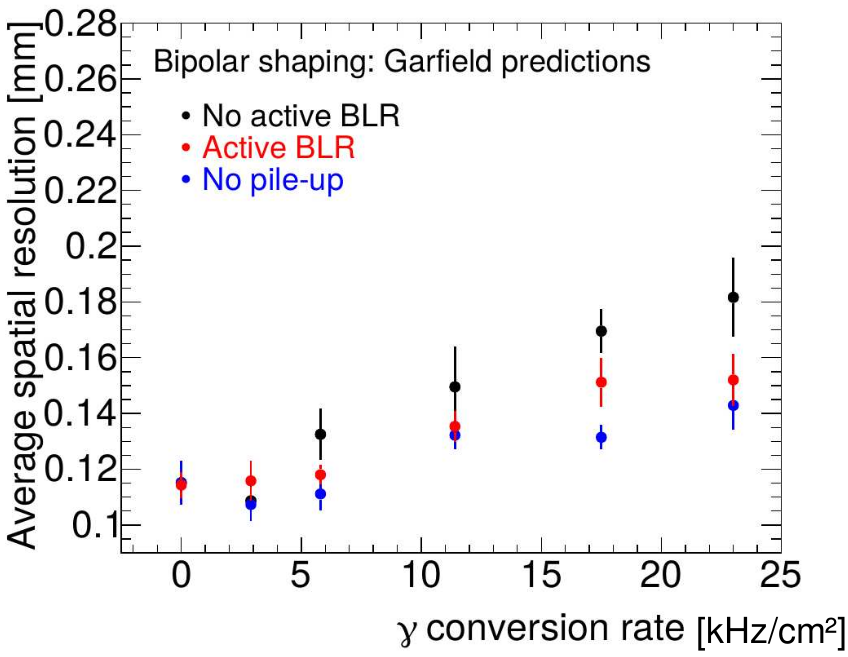}
\caption{Simulation of the degradation of the spatial resolution of an sMDT tube with increasing $\gamma$-hit rate for three scenarios. Blue: No signal pile-up, only gain drop. Black: Signal pile-up and gain drop, no active baseline restoration. Red: Signal pile-up and gain drop, perfect active baseline restoration.}
\label{Simulation}
\end{figure}

According to simulations performed on the basis of the Garfield++ framework, further improvement in spatial resolution is expected, when operating sMDT detectors with readout electronics including active baseline restoration (BLR). The result of the simulation is shown in Figure \ref{Simulation}. Electronics with BLR capability have been developed and are being tested at the GIF++ facility, with the goal of strongly reducing the remaining signal pile-up.

\section{Conclusion and Outlook}
We have shown that sMDT precision muon tracking detectors with new readout electronics are well suited for operation at the HL-LHC and future hadron colliders and an optimization of readout electronics is necessary to fully exploit the sMDT rate capability. We confirmed that faster signal shaping (as in new ASD-6 readout chip) improves the spatial resolution of sMDT drift tubes with and without background radiation. Additionally an adjustment of the operating voltage to compensate for gas gain loss at high rates due to space charge has the potential for further improvement of the spatial resolution.

The next step of rate capability improvement is the development of readout electronics with active baseline restoration. For confirmation and extension of the presented results, GIF++ testbeam measurements are being performed, using a highly energetic muon beam ($\gtrsim 100$~GeV) from the SPS. Discrete active baseline restoration circuitry is under development and is tested as well.




\end{document}